\documentclass[aps,showpacs,twocolumn,longbibliography]{revtex4-1}
\usepackage[utf8]{inputenc}
\usepackage{amsfonts}
\usepackage{amssymb}
\usepackage{amsmath}
\usepackage{graphicx}
\usepackage{epsfig}
\usepackage{subfigure}
\usepackage{appendix}
\usepackage{color}
\usepackage{hyperref}
\usepackage{fullpage}

\hypersetup{hypertex=true,
colorlinks=true,
linkcolor=blue,
anchorcolor=blue,
urlcolor=blue,
citecolor=blue}

\begin{document}

\title{Bloch oscillations in interacting systems driven by a time-dependent
magnetic field}
\author{H. P. Zhang}
\author{Z. Song}
\email{songtc@nankai.edu.cn}

\begin{abstract}
According to Faraday's law in classical physics, a varying magnetic field
stimulates an electric eddy field. Intuitively, when a classical field is
constant and imposed on a lattice, the Wannier-Stark ladders (WSL) can be
established, resulting in Bloch oscillations. In this work, we investigate
the dynamics of an interacting system on a (generalized) ring lattice
threaded by a varying magnetic flux. Based on the rigorious results, we
demonstrate that there exist many invariant subspaces in which the dynamics
is periodic when the flux varies linearly over time. Nevertheless, for a
given initial state, the evolved state differs from that driven by a linear
field. However, the probability distributions of the two states are
identical, referred to as the quantum analogue of Faraday's law.\ Our
results are ubiquitous for a wide variety of interacting systems. We
demonstrate these results through numerical simulations in an extended
fermi-Hubbard model.
\end{abstract}

\affiliation{School of Physics, Nankai University, Tianjin 300071, China}
\maketitle
\section{Introduction}

\label{Introduction}

The interplay between electromagnetic fields and quantum systems has long
been a cornerstone of modern physics, bridging the gap between classical and
quantum phenomena. One of the most fundamental principles in classical
electromagnetism, Faraday's law of induction, states that a time-varying
magnetic field induces an electric eddy field. This principle underpins a
wide range of technological applications, from electric generators to
transformers. However, its quantum counterpart remains less explored,
particularly in the context of interacting systems. Recent advancements in
the field of condensed matter physics and quantum simulation have enabled
researchers to explore novel quantum phenomena in controlled environments.
For instance, ultracold atoms in optical lattices have emerged as a powerful
platform for simulating complex quantum systems, allowing for the
manipulation of interactions and external fields with unprecedented
precision \cite{jane2003simulation,bloch2012quantum,blatt2012quantum}. This
has opened up new possibilities for investigating the dynamics of
interacting particles \cite%
{goral2002quantum,moses2015creation,moses2017new,baier2016extended,reichsollner2017quantum,Chomaz_2023}%
.

Intuitively, one might expect that a linearly varying magnetic flux can
result in Bloch oscillations (BOs) \cite%
{Bloch1929,Amar1959,Wannier1960,Waschke1993,Glueck2002} due to the constant
eddy field. In the previous work \cite{Hu2013}, the connection of two
models, an infinite tight-binding chain subjected to an arbitrary
time-dependent linear potential, and a finite ring threaded by an arbitrary
time-dependent flux, has been established for single-particle dynamics.
However, the dynamics of quantum systems under time-varying magnetic fields,
especially those with interactions between particles, present a more complex
and less understood scenario. The rigorous results in quantum many-body
systems are rare but are believed to provide valuable insights into the
characterization of dynamic behaviors within correlated systems.

In this work, we delve into the dynamics of an interacting system on a
generalized ring lattice threaded by a time-varying magnetic flux. Our
investigation is motivated by the intriguing question of how classical
principles, such as Faraday's law, manifest in the quantum realm.
Specifically, we aim to explore the periodic dynamics that emerge in
invariant subspaces when the magnetic flux varies linearly with time. We
demonstrate that, despite differences in the evolved states, the probability
distributions remain identical, a phenomenon we refer to as the quantum
analogue of Faraday's law. Our results are demonstrated in the extended fermi-Hubbard model, a versatile framework for describing interacting
fermions in lattice systems. Through numerical simulations of the
probability distributions and local currents, as functions of time, we
reveal the ubiquity of these phenomena across a wide range of interacting
systems. Our findings not only enrich the understanding of quantum dynamics
under time-varying fields but also pave the way for potential applications
in quantum control and information processing.

This paper is organized as follows. In Sec. \ref{Model and local
correspondence}, we reveal the local correspondence of general systems in
linear fields and under varying fluxes, illustrated by two small-sized
fermi-Hubbard models. Sec. \ref{Effective Wannier-Stark ladders} is
dedicated to elucidating that systems under linearly varying fluxes possess
effective Wannier-Stark ladders, which can lead to periodic dynamics. In
Sec. \ref{Bloch oscillations}, we reveal the characteristics of the quantum
Faraday's law through fermionic extended Hubbard models and observe doublon
Bloch oscillations induced by varying flux via numerical calculations.
Finally, we provide a summary in Sec. \ref{Summary}.

\section{Models and local correspondence}

\label{Model and local correspondence}

The investigations from both classical physics and modern physics imply that
there is a connection between the system subjected to a linear electric
field and the system threaded by a linearly varying flux. However, a
rigorous description has not been obtained due to two-fold obstacles. First,
the boundary conditions of the two systems are different: one is a chain,
and the other is a ring. It is impossible to establish a mapping between the
two models. Second, the interactions between particles induce a more complex
situation. In the following, we will establish this connection within the
framework of the tight-binding model.

We start with a general tight-binding model on an $N$-site chain with the
Hamiltonian in the form%
\begin{eqnarray}
H_{\mathrm{E}} &=&-\kappa \sum_{j=1}^{N-1}\sum_{\sigma =1}^{\Lambda }\left(
a_{j,\sigma }^{\dagger }a_{j+1,\sigma }+\text{\textrm{H.c}.}\right)  \notag
\\
&&+E\sum_{j=1}^{N-1}\sum_{\sigma =1}^{\Lambda }jn_{j,\sigma }+h_{\mathrm{E}},
\end{eqnarray}%
where $a_{j,\sigma }^{\dag }$\ ($a_{j,\sigma }$) is the boson or fermion
creation (annihilation) operator, with internal degree of freedom (or
flavor) $\sigma =\left[ 1,\Lambda \right] $, at the $i$th site. In the
absence of the term $h_{\mathrm{E}}$, $H_{\mathrm{E}}$\ describes a chain
with uniform hopping strength $\kappa $, and a linear potential with slope $%
E $. Here $h_{\mathrm{E}}=h_{\mathrm{E}}(\left\{ n_{l,\sigma _{l}}\right\} )$
is a term describing local particle-particle interactions and local
potentials. It is an operator consisting of a set of particle number
operators $(n_{1,\sigma _{1}},...,n_{l,\sigma _{l}},...,n_{N,\sigma _{N}})$.
In order to investigate the solution of the Schrodinger equation%
\begin{equation}
i\frac{\partial }{\partial t}|\psi _{\mathrm{E}}(t)\rangle =H_{\mathrm{E}%
}|\psi _{\mathrm{E}}(t)\rangle ,
\end{equation}%
one can take the rotating frame by introducing the transformation \cite%
{Bukov2015} 
\begin{equation}
V(t)=\exp (iEt\sum_{j,\sigma }jn_{j,\sigma }).  \label{Vt}
\end{equation}%
For both boson and fermion operators, we always have 
\begin{equation}
V(t)a_{j,\sigma }V^{-1}(t)=e^{-iEtj}a_{j,\sigma },
\end{equation}%
and%
\begin{equation}
V(t)h_{\mathrm{E}}V^{-1}(t)=h_{\mathrm{E}},
\end{equation}%
which result in the Schrodinger equation 
\begin{equation}
H_{\mathrm{ER}}|\psi _{\mathrm{ER}}(t)\rangle =i\frac{\partial }{\partial t}%
|\psi _{\mathrm{ER}}(t)\rangle ,
\end{equation}%
in the rotating frame with $|\psi _{\mathrm{ER}}(t)\rangle =V(t)|\psi _{%
\mathrm{E}}(t)\rangle $. The corresponding Hamiltonian has the from%
\begin{equation}
H_{\mathrm{ER}}=-\kappa \sum_{j=1}^{N-1}\sum_{\sigma =1}^{\Lambda }\left(
e^{-iEt}a_{j,\sigma }^{\dagger }a_{j+1,\sigma }+\text{\textrm{H.c}.}\right)
+h_{\mathrm{E}}.
\end{equation}%
Comparing the Hamiltonian $H_{\mathrm{ER}}$\ with the original one $H_{%
\mathrm{E}}$, we note that the linear potential term is replaced by the
phase factor in the hopping term. This inspires us to consider another
Hamiltonian with complex hopping strength arising from the magnetic flux.

Now, we consider a similar tight-binding model on an $M$-site ring with a
time-dependent magnetic flux, $\mathrm{\Phi }(t)$\ threaded through it. The
Hamiltonian has the form

\begin{equation}
H_{\mathrm{\Phi }}=-\kappa \sum_{j=1}^{M}\sum_{\sigma =1}^{\Lambda }\left(
e^{-i\mathrm{\Phi }(t)/M}a_{j,\sigma }^{\dagger }a_{j+1,\sigma }+\text{%
\textrm{H.c}.}\right) +h_{\mathrm{\Phi }},
\end{equation}%
where the periodic boundary condition $a_{M+1,\sigma }=a_{1,\sigma }$\ is
taken. Here the term $h_{\mathrm{\Phi }}$\ is similar to $h_{\mathrm{E}}$,
describing local particle-particle interactions and local potentials. We
would like to point out that we deliberately do not take $M=N$, and do not
assume completely identical $h_{\mathrm{\Phi }}$\ and $h_{\mathrm{E}}$,
without losing generality.

Obviously, we cannot conclude that the two Hamiltonians $H_{\mathrm{\Phi }}$%
\ and $H_{\mathrm{ER}}$\ are equivalent even when taking $M=N$, $\mathrm{%
\Phi }(t)/M=Et$, and $h_{\mathrm{\Phi }}=h_{\mathrm{E}}$. However, we can
establish the following local equivalence between them. Considering that the
two Hamiltonians satisfy $h_{\mathrm{\Phi }}=h_{\mathrm{E}}$\ within a real
space region $j\in \left[ l_{\text{\textrm{L}}},l_{\text{\textrm{R}}}\right] 
$, if a given initial state $|\phi (0)\rangle $ and its evolved state are
local states in this region, we have%
\begin{eqnarray}
|\phi (t)\rangle &=&\mathcal{T}e^{-i\int_{0}^{t}H_{\mathrm{ER}}(t^{\prime })%
\mathrm{d}t^{\prime }}|\phi (0)\rangle \\
&=&|\psi _{\mathrm{\Phi }}(t)\rangle =\mathcal{T}e^{-i\int_{0}^{t}H_{\mathrm{%
\Phi }}(t^{\prime })\mathrm{d}t^{\prime }}|\phi (0)\rangle ,  \notag
\end{eqnarray}%
where $\mathcal{T}$\ is the time-order operator. It indicates that the two
evolved states, driven by the respective Hamiltonians, exhibit identical
dynamics. The proof is straightforward since the local state always has no
particle probability beyond the given region. That is, $\left\vert
a_{j,\sigma }|\phi (t)\rangle \right\vert ^{2}=0$\ for $j\notin %
\left[ l_{\text{\textrm{L}}},l_{\text{\textrm{R}}}\right] $. Accordingly, we
have 
\begin{equation}
V(t)e^{-iH_{\mathrm{E}}t}|\phi (0)\rangle =|\psi _{\mathrm{\Phi }}(t)\rangle
,
\end{equation}%
which indicates that evolved states under the two Hamiltonians $H_{\mathrm{%
\Phi }}$\ and $H_{\mathrm{E}}$\ are not identical but are connected by a
mapping. This is referred to as local correspondence of the two systems.\
This can also be regarded as a quantum version of Faraday's law, which
establishes the relation between a varying magnetic flux and a local field.

\begin{figure}[t]
\centering
\includegraphics[width=0.9\linewidth]{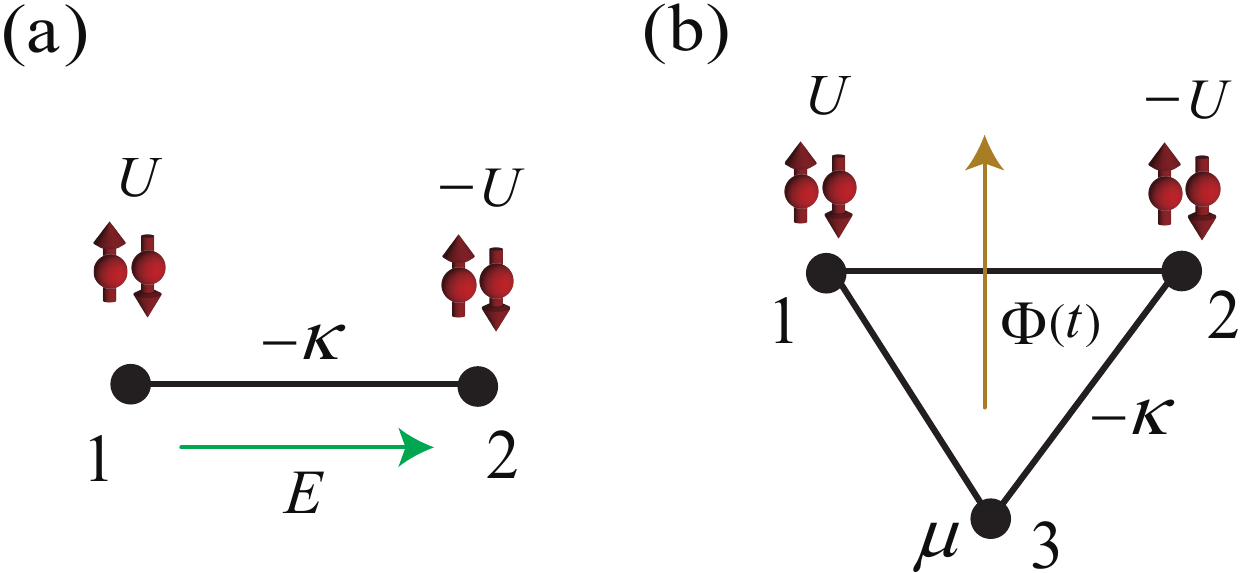}  
\caption{Schematic illustrations of the Hamiltonian in Eqs. ({\protect\ref%
{H1}}) and ({\protect\ref{H2}}), which represent (a) a two-site chain in an
electric field $E=U$ and (b) a three-site ring with a magnetic flux $\Phi=3Ut$
. Hopping strength is $-\protect\kappa$. On-site interactions at the 1st and
2nd sites are $U$ and $-U$, respectively. Chemical potential at the 3rd site
is $\protect\mu$, which plays the role of confining the particles within the
dimer, resulting in local correspondence. The dynamics of the two systems
become more similar as $\protect\mu$ increases.}
\label{fig1}
\end{figure}
\begin{figure}[t]
\centering
\includegraphics[width=0.95\linewidth]{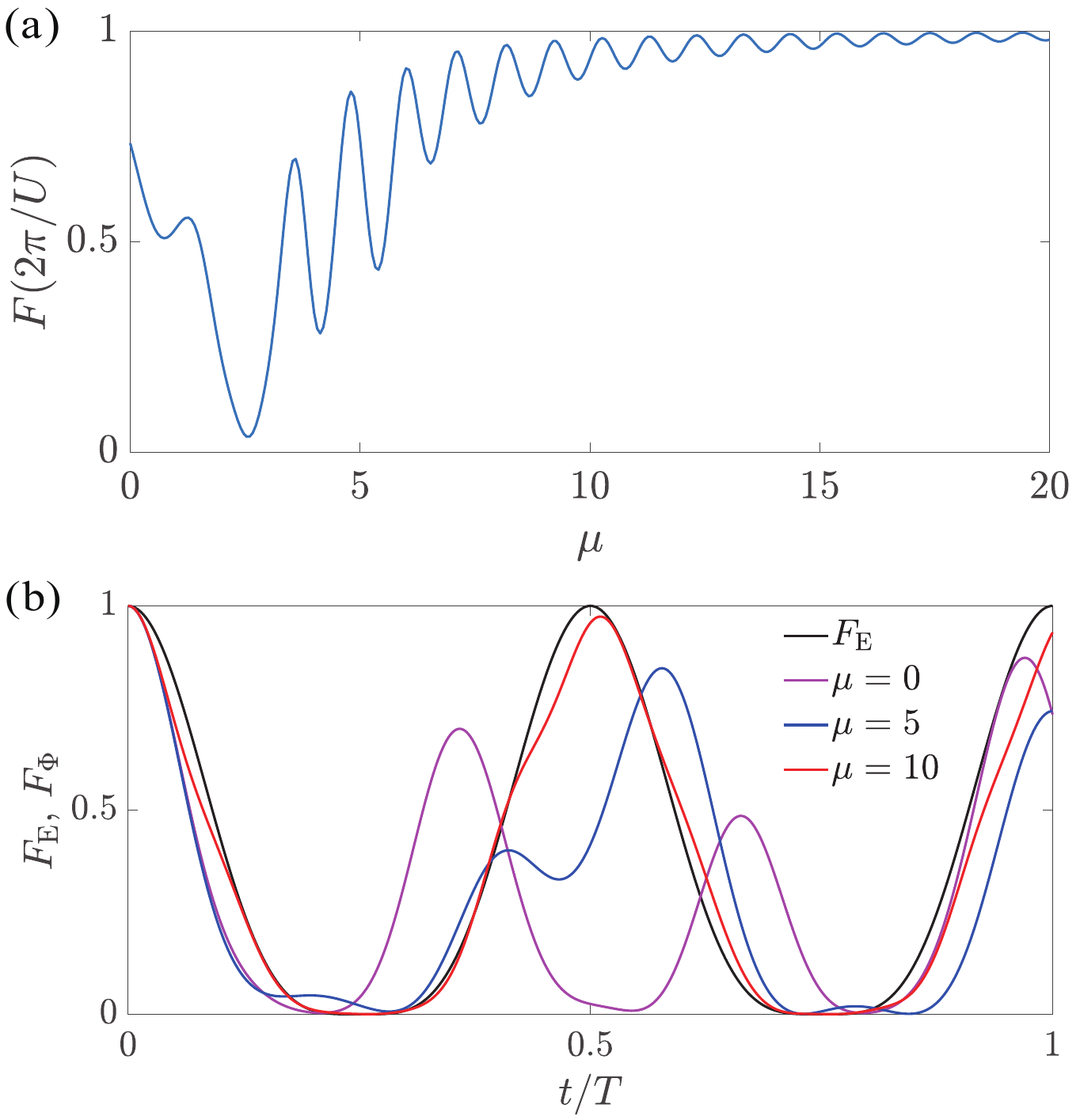}
\caption{Plots of three types of fidelity defined in Eqs. (\protect\ref{Ft}%
), (\protect\ref{F_Et}) and (\protect\ref{F_Phit}), which describe the
dynamical similarity of the two systems in Fig. \protect\ref{fig1} for
different values of $\protect\mu$. (a) Fidelity between the evolved states
from the same initial state in Fig.\ \protect\ref{fig1}(a) and Fig.\ \protect
\ref{fig1}(b) with different $\protect\mu$, at time $t=2\protect\pi/U$. (b)
Fidelities $F_\mathrm{E}$ in Fig.\ \protect\ref{fig1}(a) and $F_{\Phi}$
Fig.\ \protect\ref{fig1}(b) for different $\protect\mu$. It can be seen that
as $\protect\mu$ increases, (a) $F$ approaches 1, and (b) the curves of $F_%
\mathrm{E}$ and $F_{\Phi}$ coincide, indicating the local correspondence in
the large $\protect\mu$ limit. Here, $\protect\kappa=U=1$, $T=2\protect\pi/U$%
.}
\label{fig2}
\end{figure}

Before further investigations into the local correspondence, we now consider
two small-sized fermi-Hubbard models to demonstrate the results. In
these models, the flavor $\Lambda =2$, with $a_{j,1}=c_{j,\uparrow }$\ and $%
a_{j,2}=c_{j,\downarrow }$ being the fermion operators. The Hamiltonians are
given by%
\begin{eqnarray}
H_{1} &=&-\kappa \sum_{\sigma =\uparrow ,\downarrow }(c_{1,\sigma }^{\dagger
}c_{2,\sigma }+\text{\textrm{H.c}.})+U\sum_{\sigma =\uparrow ,\downarrow
}n_{2,\sigma }  \notag \\
&&+U(n_{1,\uparrow }n_{1,\downarrow }-n_{2,\uparrow }n_{2,\downarrow }),
\label{H1}
\end{eqnarray}%
and%
\begin{eqnarray}
H_{2} &=&-\kappa \sum_{\sigma =\uparrow ,\downarrow }(e^{-iUt}c_{1,\sigma
}^{\dagger }c_{2,\sigma }+e^{-iUt}c_{2,\sigma }^{\dagger }c_{3,\sigma } 
\notag \\
&&+e^{-iUt}c_{3,\sigma }^{\dagger }c_{1,\sigma })+\text{\textrm{H.c}.} \\
&&+U(n_{1,\uparrow }n_{1,\downarrow }-n_{2,\uparrow }n_{2,\downarrow })+\mu
\sum_{\sigma =\uparrow ,\downarrow }n_{3,\sigma },  \notag  \label{H2}
\end{eqnarray}%
where $c_{j,\sigma }^{\dag }$\ ($c_{j,\sigma }$) is the fermion creation
(annihilation) operator, with spin index $\sigma =\uparrow ,\downarrow $.
The Hamiltonian $H_{1}$ describes a Hubbard dimer with resonant on-site
interaction strength and a linear potential $U$. The Hamiltonian $H_{2}$
describes a $3$-site Hubbard ring threading a resonantly varying magnetic
flux. There is an on-site potential $\mu $ at the third site. Fig. \ref{fig1}
is the schematic diagram of the two systems.

According to our above analysis, the evolved states confined within the
dimers of the two systems are connected by a mapping. In the Appendix, we
derive the matrix representations of the two Hamiltonians in the $2$-fermion
invariant subspaces, and the corresponding derivations are provided.
Obviously, the 3rd-site can be separated from the dimer when taking $\mu
\rightarrow \infty $. Then the effective Hamiltonian of the dimer becomes%
\begin{eqnarray}
H_{2}^{\text{\textrm{eff}}} &=&-\kappa \sum_{\sigma =\uparrow ,\downarrow
}e^{-iUt}c_{1,\sigma }^{\dagger }c_{2,\sigma }+\text{\textrm{H.c}.}  \notag
\\
&&+U(n_{1,\uparrow }n_{1,\downarrow }-n_{2,\uparrow }n_{2,\downarrow }),
\end{eqnarray}%
which is nothing but the expression of $H_{1}$\ in the rotating frame.
According to the above analysis, the potential $\mu $ plays the role of
confining the particles within the dimer, resulting in local correspondence.
For finite $\mu $, the efficiency of this correspondence is determined by
the value of $\mu $. To demonstrate this point, we numerically compute the
time evolution of an initial state $|\phi (0)\rangle =c_{1,\uparrow }^{\dag
}c_{1,\downarrow }^{\dag }\left\vert 0\right\rangle $\ under the two
Hamiltonians $H_{1}$\ and $H_{2}$. We employ the fidelities $F(t)$, $F_{%
\mathrm{E}}(t)$,\ and $F_{\mathrm{\Phi }}(t)$, given in the Appendix, to
characterize the similarity of the two evolved states for difference values
of $\mu $. We plot $F(2\pi /U)$ as function of $\mu $, $F_{\mathrm{E}}(t)$\
and $F_{\mathrm{\Phi }}(t)$, as function of time\ in the Fig. \ref{fig2}.

We can see that $F_{\mathrm{E}}(t)$\ is exactly periodic with a period of $%
\pi /\kappa $. The plot of $F_{\mathrm{\Phi }}(t)$ for several typical
values of $\mu $ shows that $F_{\mathrm{\Phi }}(t)$ approaches $F_{\mathrm{E}%
}(t)$\ as $\mu $\ increases, indicating the local correspondence in the
large $\mu$ limit.

\begin{figure}[t]
\centering
\includegraphics[width=0.95\linewidth]{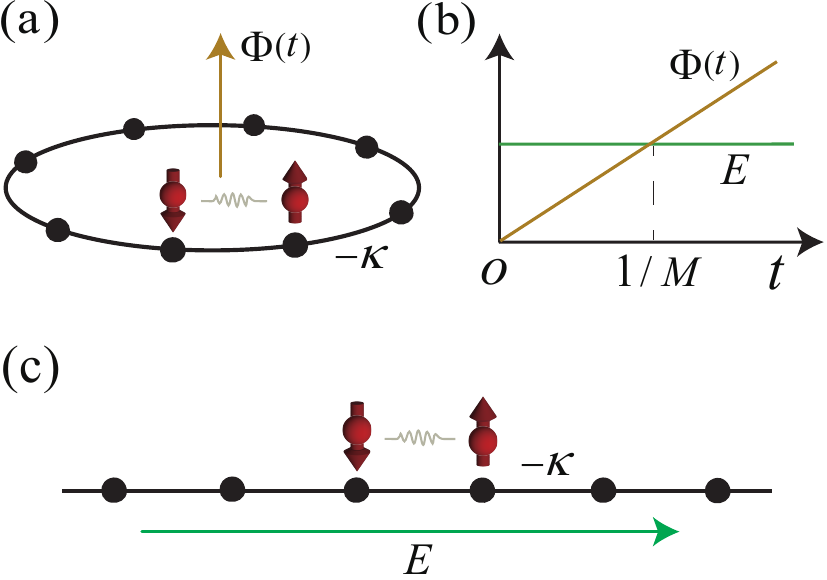}
\caption{Schematic diagrams of (a) an $M$-site ring with a varying magnetic flux $%
\Phi(t)=MEt$ and (c) an $N$-site chain with an external electric field $E$, both
featuring the same interaction terms. When $M$ and $N$ are sufficiently large, there is a local correspondence
between the two systems, i.e., $V(t)|\protect\psi _{\mathrm{E}}(t)\rangle =|%
\protect\psi _{\mathrm{\Phi}}(t)\rangle$, which leads to the same local
particle density $\langle n_j \rangle$ but different local dimensionless
currents $\langle J_j \rangle$. The energy levels of Hamiltonian (c) consist
of multiple sets of WSLs and exhibit periodic dynamics, which leads to Bloch
oscillations driven by a time-dependent magnetic field, as shown in Fig. 
\protect\ref{fig6}.}
\label{fig3}
\end{figure}

\section{Effective Wannier-Stark ladders}

\label{Effective Wannier-Stark ladders}

\begin{figure*}[t]
\centering
\includegraphics[width=0.95\linewidth]{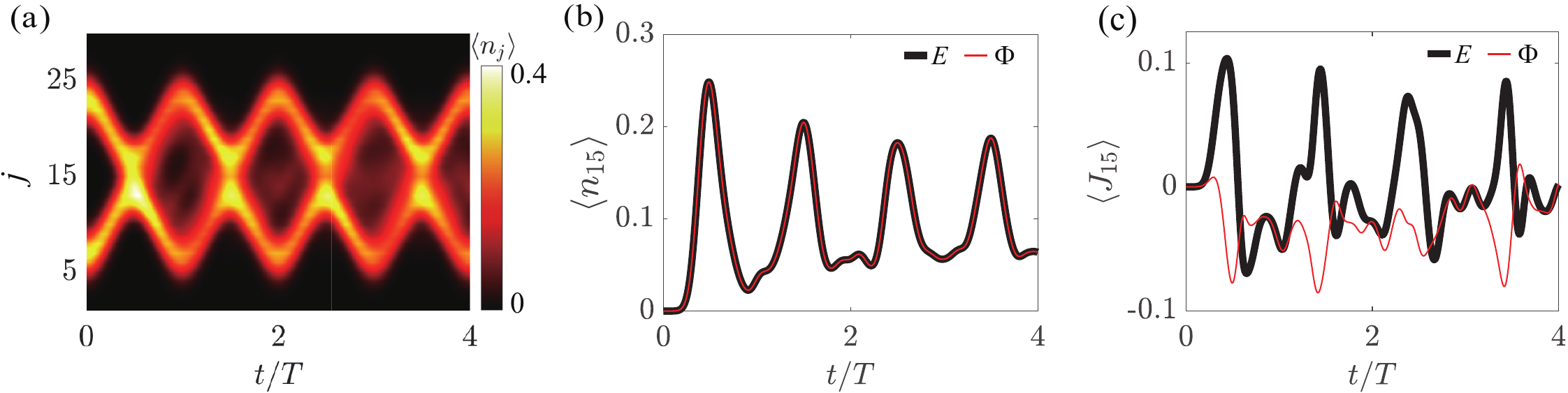}  
\caption{Plots of $\langle n_j \rangle$ and $\langle J_j \rangle$ defined in
Eqs. (\protect\ref{n_j}) and (\protect\ref{J_j}), obtained by numerical
diagonalization for the separated Gaussian initial state $|\protect\phi %
_{1}(0)\rangle$ in Hamiltonian $H_\mathrm{E}$ and $H_{\Phi}$ defined in
Eqs.\ (\protect\ref{H_E}) and (\protect\ref{H_Phi}), repectively. (a) The
evolution diagrams of $\langle n_j \rangle_\mathrm{E}$ and $\langle n_j
\rangle_\mathrm{\Phi}$ are identical, which depict the collision of two
single-particle Gaussian wave packets. For the 15-th site, (b) and (c) show
that the evolved states under $H_\mathrm{E}$ and $H_{\Phi}$ have the same
local particle density but different local dimensionless currents, which
indicate the characteristics of the quantum Faraday's law. Here, $\protect%
\kappa=U=V=1$, $E=0.6$, $T=2\protect\pi/E$. The factor of Gaussian
wavepactet $\protect\alpha$ equals $0.1$.}
\label{fig4}
\end{figure*}
\begin{figure*}[t]
\centering
\includegraphics[width=0.95\linewidth]{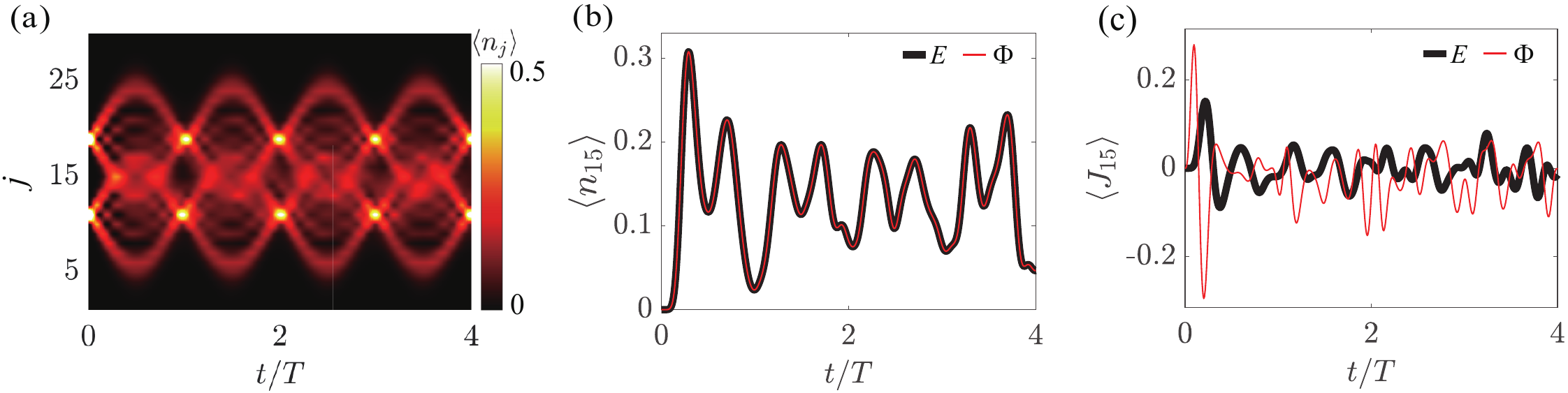}  
\caption{The same plots as Fig. \protect\ref{fig4} for the case where the
initial state $|\protect\phi _{2}(0)\rangle$ is two separated site-states
with opposite spins, given by Eq.\ (\protect\ref{phi2}).}
\label{fig5}
\end{figure*}
\begin{figure*}[t]
\centering
\includegraphics[width=0.95\linewidth]{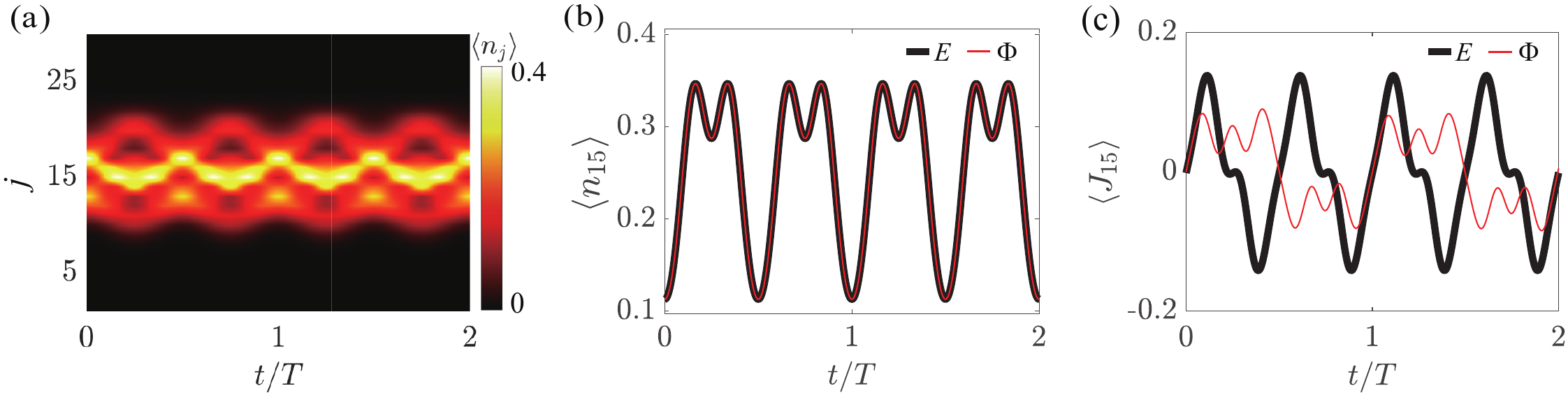}  
\caption{The same plots as Fig. \protect\ref{fig4} for the case where the
initial state $|\protect\phi _{3}(0)\rangle$ is a superposition of localized
eigenstates $|\protect\psi _{m}\rangle $ of $H_\mathrm{E}$ in a set of WSL,
given by Eq.\ (\protect\ref{phi3}). It can be seen that the oscillation
periods of $\langle n_j \rangle_{\mathrm{E},\Phi}$ and $\langle J_j
\rangle_{ \mathrm{E}}$ are $T/2$, while that of $\langle J_j \rangle_{\Phi}$
is $T$, where $T=2\protect\pi/E$ is the period of single-particle Bloch
oscillations. This results from the period of the local correspondence
mapping $V(t)$ being $T$. Here, $|\protect\psi _{m}\rangle $ is chosen as
the eigenstate with the largest $\langle n_m \rangle$, which can be
identified through numerical calculations. The factor of wavepactet $\protect%
\alpha$ equals $0.4$.}
\label{fig6}
\end{figure*}

In this section, we will extend our investigation to the quantum Faraday's
law in terms of its application aspect. We consider the Hamiltonians $H_{%
\mathrm{E}}$\ and $H_{\mathrm{\Phi }}$\ with on an infinite size lattice, in
which the interacting terms have translational symmetry and $h_{\mathrm{E}%
}=h_{\mathrm{\Phi }}$. Fig. \ref{fig3} illustrates the two types of
Hamiltonians. Before proceeding, we would like to give a brief review of the
features of $H_{\mathrm{E}}$ in the following form 
\begin{eqnarray}
H_{\mathrm{E}} &=&-\kappa \sum_{j=-\infty }^{\infty }\sum_{\sigma
=1}^{\Lambda }\left( a_{j,\sigma }^{\dagger }a_{j+1,\sigma }+\text{\textrm{%
H.c}.}\right)  \notag \\
&&+E\sum_{j=-\infty }^{\infty }\sum_{\sigma =1}^{\Lambda }jn_{j,\sigma }+h_{%
\mathrm{E}},
\end{eqnarray}%
where $h_{\mathrm{E}}$\ satisfies following conditions:

(i) The total particle number $\sum_{j=-\infty }^{\infty }\sum_{\sigma
=1}^{\Lambda }a_{j,\sigma }^{\dag }a_{j,\sigma }$\ is conservative, that is%
\begin{equation}
\lbrack \sum_{j=-\infty }^{\infty }\sum_{\sigma =1}^{\Lambda }a_{j,\sigma
}^{\dag }a_{j,\sigma },h_{\mathrm{E}}]=0.
\end{equation}

(ii) Hamiltonian $h_{\mathrm{E}}$\ has translational symmetry, that is%
\begin{equation}
\left[ T_{r},h_{\mathrm{E}}\right] =0,
\end{equation}%
where $T_{r}$\ is the translational operator defined as 
\begin{equation}
T_{r}a_{j,\sigma }T_{r}^{-1}=a_{j+r,\sigma }.
\end{equation}%
According to the theorem proposed in the Ref. \cite{Zhang2025}, the energy
levels of Hamiltonian $H_{\mathrm{E}}$ must consist of multiple sets of WSLs
with an identical real level spacing $nrE$, which is independent of the
details of $h_{\mathrm{E}}$. Here, $n$ is the particle number of the
localized eigenstates. Consequently, there exist multiple sets of localized
initial states, which exhibit periodic dynamics with a period of $2\pi
/(nrE) $. This conclusion also holds for the corresponding $H_{\mathrm{ER}}$%
, which is expressed in the following form%
\begin{equation}
H_{\mathrm{ER}}=-\kappa \sum_{j=-\infty }^{\infty }\sum_{\sigma =1}^{\Lambda
}\left( e^{-iEt}a_{j,\sigma }^{\dagger }a_{j+1,\sigma }+\text{\textrm{H.c}.}%
\right) +h_{\mathrm{E}}.
\end{equation}%
Furthermore, based on the local correspondence proposed in the above
section, the same conclusion can be extended to the corresponding
time-dependent Hamiltonian on a ring lattice%
\begin{equation}
H_{\mathrm{\Phi }}=-\kappa \sum_{j=1}^{M}\sum_{\sigma =1}^{\Lambda }\left(
e^{-iEt}a_{j,\sigma }^{\dagger }a_{j+1,\sigma }+\text{\textrm{H.c}.}\right)
+h_{\mathrm{E}},
\end{equation}%
where $M$ is sufficiently large. In this sense, there also exist multiple
sets of localized initial states for $H_{\mathrm{\Phi }}$, which exhibit
periodic dynamics with a period of $2\pi /(nrE)$. Then, the Hamiltonian $H_{%
\mathrm{\Phi }}$\ describes a system that possesses multiple sets of\
effective Wannier-Stark ladders. However, it is worth noting that for a
given localized initial state, the two evolved states under the two
Hamiltonians $H_{\mathrm{E}}$\ and\ $H_{\mathrm{\Phi }}$ are not exactly
identical due to the mapping $V(t)$\ between them. Such differences can be
measured by certain observables, which will be discussed in the following
section.

\section{Bloch oscillations}

\label{Bloch oscillations}

In this section, we will demonstrate the obtained results in an extended
fermi-Hubbard model, where the flavor $\Lambda =2$, with $%
a_{j,1}=c_{j,\uparrow }$\ and $a_{j,2}=c_{j,\downarrow }$ being the fermion
operators. The corresponding two Hamiltonians are%
\begin{eqnarray}
H_{\mathrm{E}} &=&-\kappa \sum_{j=-\infty }^{\infty }\sum_{\sigma =\uparrow
,\downarrow }c_{j,\sigma }^{\dagger }c_{j+1,\sigma }+\text{\textrm{H.c}.}%
+E\sum_{j,\sigma }jn_{j,\sigma }  \notag \\
&&+\sum_{j=-\infty }^{\infty }(Un_{j,\uparrow }n_{j,\downarrow
}+Vn_{j}n_{j+1}),  \label{H_E}
\end{eqnarray}%
and

\begin{eqnarray}
H_{\mathrm{\Phi }} &=&-\kappa \sum_{j=-\infty }^{\infty }\sum_{\sigma
=\uparrow ,\downarrow }e^{-iEt}c_{j,\sigma }^{\dagger }c_{j+1,\sigma }+\text{%
\textrm{H.c}.}  \notag \\
&&+\sum_{j=-\infty }^{\infty }(Un_{j,\uparrow }n_{j,\downarrow
}+Vn_{j}n_{j+1}),  \label{H_Phi}
\end{eqnarray}%
respectively. Here, $n_{j}=n_{j,\uparrow }+n_{j,\downarrow }$\ is the total
fermion number operator at the $j$-th site. The particle-particle
interaction includes on-site and nearest-neighbouring (NN) interactions.

Considering an initial state $|\phi (0)\rangle $, its evoved state $|\phi _{%
\mathrm{E}}(t)\rangle $\ can be expressed in the Fock as follows%
\begin{equation}
|\phi _{\mathrm{E}}(t)\rangle =\sum_{\left\{ n_{l,\sigma _{l}}\right\}
}C_{\left\{ n_{l,\sigma _{l}}\right\} }(t)\left\vert \left\{ n_{l,\sigma
_{l}}\right\} \right\rangle ,
\end{equation}%
where $\left\vert \left\{ n_{l,\sigma _{l}}\right\} \right\rangle
=\prod_{\left\{ n_{l,\sigma _{l}}\right\} }c_{l,\sigma _{l}}^{\dag
}\left\vert 0\right\rangle $ denotes the basis of the Fock space, with $%
\left\vert 0\right\rangle $\ being the vacuum state. Accordingly, the evoved
state under the Hamiltonian\ $H_{\mathrm{\Phi }}$\ can be expressed as%
\begin{equation}
|\phi _{\mathrm{\Phi }}(t)\rangle =\sum_{\left\{ n_{l,\sigma _{l}}\right\}
}C_{\left\{ n_{l,\sigma _{l}}\right\} }(t)D_{\left\{ n_{l,\sigma
_{l}}\right\} }\left\vert \left\{ n_{l,\sigma _{l}}\right\} \right\rangle ,
\end{equation}%
with the factor%
\begin{equation}
D_{\left\{ n_{l,\sigma _{l}}\right\} }=\exp \left( i\sum_{n_{s,\sigma
_{s}}\in \left\{ n_{l,\sigma _{l}}\right\} }sn_{s,\sigma _{s}}Et\right) .
\end{equation}%
In order to demonstrate the similarity and difference between the two states 
$|\phi _{\mathrm{E}}(t)\rangle $\ and $|\phi _{\mathrm{\Phi }}(t)\rangle $,
we consider the expectation values of two observables. One is the local
particle density, while the other is the local dimensionless current.
Straightforward derivations show that

\begin{eqnarray}
&&\left\langle n_{j}\right\rangle _{\mathrm{E}}=\sum_{\sigma =\uparrow
,\downarrow }\langle \psi _{\mathrm{E}}(t)|n_{j,\sigma }|\psi _{\mathrm{E}%
}(t)\rangle  \notag \\
&=&\sum_{\sigma =\uparrow ,\downarrow }\langle \psi _{\mathrm{\Phi }%
}(t)|n_{j,\sigma }|\psi _{\mathrm{\Phi }}(t)\rangle =\left\langle
n_{j}\right\rangle _{\mathrm{\Phi }},  \label{n_j}
\end{eqnarray}%
and%
\begin{eqnarray}  \label{J_j}
&&\left\langle J_{j}\right\rangle _{\mathrm{E}}=\sum_{\sigma =\uparrow
,\downarrow }i\langle \psi _{\mathrm{E}}(t)|(c_{j,\sigma }^{\dag
}c_{j+1,\sigma }-\text{\textrm{H.c}.})|\psi _{\mathrm{E}}(t)\rangle  \notag
\\
&=&\sum_{\sigma =\uparrow ,\downarrow }i\langle \psi _{\mathrm{\Phi }%
}(t)|(e^{-iEt}c_{j,\sigma }^{\dag }c_{j+1,\sigma }-\text{\textrm{H.c}.}%
)|\psi _{\mathrm{\Phi }}(t)\rangle  \notag \\
&\neq &\sum_{\sigma =\uparrow ,\downarrow }i\langle \psi _{\mathrm{\Phi }%
}(t)|(c_{j,\sigma }^{\dag }c_{j+1,\sigma }-\text{\textrm{H.c}.})|\psi _{%
\mathrm{\Phi }}(t)\rangle =\left\langle J_{j}\right\rangle _{\mathrm{\Phi }},
\notag \\
\end{eqnarray}%
which indicate the characteristics of the quantum Faraday's law.

To verify and demonstrate the above analysis, numerical simulations are
performed to investigate the dynamic behaviors driven by the two fermionic
Hamiltonians $H_{\mathrm{E}}$ and $H_{\mathrm{\Phi }}$, as given above.\ We
compute the temporal evolution for three types of initial states: (i) Two
separated Gaussian wavepacket states with opposite spins. The initial state
is expressed\ in the form%
\begin{eqnarray}
|\phi _{1}(0)\rangle &=&\sqrt{\frac{2\alpha ^{2}}{\pi }}\left(
\sum_{j}e^{-\alpha ^{2}(j-j_{A})^{2}}e^{-i\frac{\pi }{2}j}c_{j,\uparrow
}^{\dag }\right)  \notag \\
&&\times \left( \sum_{j}e^{-\alpha ^{2}(j-j_{B})^{2}}e^{i\frac{\pi }{2}%
j}c_{j,\downarrow }^{\dag }\right) |0\rangle ,  \label{phi1}
\end{eqnarray}%
\ which is the product state of two wavepackets centered at $j_{A}$-th and $%
j_{B}$-th sites, with group velocities $\pm 2\kappa $, respectively. The
profile of the wavepackets is determined by the factor$\ \alpha $. (ii) Two
separated site-states with opposite spins. The initial state is expressed\
in the form%
\begin{equation}
|\phi _{2}(0)\rangle =c_{j_{A},\uparrow }^{\dag }c_{j_{B},\downarrow }^{\dag
}|0\rangle ,  \label{phi2}
\end{equation}%
which is the product state of two site-states at $j_{A}$-th and $j_{B}$-th
sites, respectively. (iii) A superposition of localized eigenstates of $H_{%
\mathrm{E}}$\ in a set of WSL. The initial state is expressed\ in the form%
\begin{equation}
|\phi _{3}(0)\rangle =\left( \frac{2\alpha ^{2}}{\pi }\right) ^{\frac{1}{4}%
}\left( \sum_{m}e^{-\alpha ^{2}(m-m_{c})^{2}}|\psi _{m}\rangle \right) .
\label{phi3}
\end{equation}%
where $|\psi _{m}\rangle $\ is the eigenstate of a set of WSL in $H_{\mathrm{%
E}}$\ with energy level spacings of $2E$, and $|\psi _{m_{c}}\rangle $\ is
the eigenstate with the largest proportion. It can be shown that $|\psi
_{m+1}\rangle =T_{1}|\psi _{m}\rangle $, where $T_{1}$\ is the translational
operator defined as $T_{1}c_{j,\sigma }T_{1}^{-1}=c_{j+1,\sigma }.$

Based on the numerical simulations on the evolved state for the initial
states $|\phi _{i}(0)\rangle $\ ($i=1,2,3$) under the two fermionic
Hamiltonians $H_{\mathrm{E}}$ and $H_{\mathrm{\Phi }}$, on finite lattice,\
we compute the corresponding quantities $\left\langle n_{j}\right\rangle _{%
\mathrm{E}}$, $\left\langle n_{j}\right\rangle _{\mathrm{\Phi }}$, $%
\left\langle J_{j}\right\rangle _{\mathrm{E}}$\ and $\left\langle
J_{j}\right\rangle _{\mathrm{\Phi }}$, respectively. We plot these
quantities in Figs. \ref{fig4}, \ref{fig5} and \ref{fig6}. {These numerical
results accord with our above analysis}: (i) For initial states $|\phi
_{1}(0)\rangle $\ and $|\phi _{2}(0)\rangle $, at beginning, the dynamics is
single-particle Bloch oscillations, since the interactions between two
particles have no effect. When two particle collide, the interactions
between two particles switch on, resulting in quasi periodic dynamic
behaviors. In addition, the results evidently indicate the relations, $%
\left\langle n_{j}\right\rangle _{\mathrm{E}}=\left\langle
n_{j}\right\rangle _{\mathrm{\Phi }}$ and $\left\langle J_{j}\right\rangle _{%
\mathrm{E}}\neq \left\langle J_{j}\right\rangle _{\mathrm{\Phi }}$; (ii) For
initial state $|\phi _{3}(0)\rangle $, the dynamics exhibits periodic
dynamic behaviors. As expected, it is also observed that $\left\langle
n_{j}\right\rangle _{\mathrm{E}}=\left\langle n_{j}\right\rangle _{\mathrm{%
\Phi }}$ and $\left\langle J_{j}\right\rangle _{\mathrm{E}}\neq \left\langle
J_{j}\right\rangle _{\mathrm{\Phi }}$.

\section{Summary}

\label{Summary}

In summary, we have revealed the local correspondence of interacting systems
under linear potentials and linearly varying magnetic flux, which
indicates that the evolved states under the two are not identical but are
connected by a mapping. This leads to the same local particle density but
different local dimensionless currents, which are characteristics of the
quantum Faraday's law. We also demonstrate that there exist multiple sets of
localized initial states for $H_{\mathrm{\Phi }}$, which exhibit periodic
dynamics with a period of $2\pi /(nrE)$. To verify this, we investigate
fermionic extended Hubbard models, numerically compute the evolution of
three typical initial states under $H_{\mathrm{E}}$\ and $H_{\mathrm{\Phi }}$%
, and observe doublon Bloch oscillations induced by varying flux. These
findings bridge the gap between electric and magnetic fields at the quantum
level and lay the foundation for exploring novel physical phenomena induced
by magnetic fields in interacting systems.

\acknowledgments This work was supported by NSFC (Grant No.12374461).

\section*{Appendix}

\label{Appendix}

In this appendix, we present the explicit forms of the matrix
representations of the Hamiltonians $H_{1}$\ and $H_{2}$, given in Eqs. (\ref%
{H1}) and (\ref{H2}), respectively. We also investigate the dynamic
behaviors driven by the two Hamiltonians. We focus on the invariant subspace
with zero spin and two fermions. The subspace is spanned by the following
basis

\begin{equation}
\left( c_{1,\uparrow }^{\dag }c_{1,\downarrow }^{\dag },d_{12}^{\dag
},c_{2,\uparrow }^{\dag }c_{2,\downarrow }^{\dag },d_{13}^{\dag
},d_{23}^{\dag },c_{3,\uparrow }^{\dag }c_{3,\downarrow }^{\dag }\right)
\left\vert 0\right\rangle ,  \label{basis}
\end{equation}%
where $d_{ij}^{\dag }=(c_{i,\uparrow }^{\dag }c_{j,\downarrow }^{\dag
}-c_{i,\downarrow }^{\dag }c_{j,\uparrow }^{\dag })/\sqrt{2}$ is singlet
pair operator across two sites $i$ and $j$.

The matrix representations of the Hamiltonians $H_{1}$\ and $H_{2}$ in the
above basis set are both $6\times 6$\ matrices. However, the matrix of $%
H_{1} $\ can be reduced to $3\times 3$\ matrix%
\begin{equation}
h_{1}=-\sqrt{2}\kappa \left( 
\begin{array}{ccc}
0 & 1 & 0 \\ 
1 & 0 & 1 \\ 
0 & 1 & 0%
\end{array}%
\right) +U.
\end{equation}%
The matrix of $H_{2}$\ is

\begin{eqnarray}
h_{2} &=&\left( 
\begin{array}{cccccc}
-\frac{U}{\sqrt{2}\kappa } & e^{-iUt} & 0 & e^{iUt} & 0 & 0 \\ 
e^{iUt} & 0 & e^{-iUt} & \frac{e^{-iUt}}{\sqrt{2}} & e^{iUt} & 0 \\ 
0 & e^{iUt} & \frac{U}{\sqrt{2}\kappa } & 0 & e^{-iUt} & 0 \\ 
e^{-iUt} & \frac{e^{iUt}}{\sqrt{2}} & 0 & -\frac{\mu }{\sqrt{2}\kappa } & 
\frac{e^{-iUt}}{\sqrt{2}} & e^{iUt} \\ 
0 & e^{-iUt} & e^{iUt} & \frac{e^{iUt}}{\sqrt{2}} & -\frac{\mu }{\sqrt{2}%
\kappa } & e^{-iUt} \\ 
0 & 0 & 0 & e^{-iUt} & e^{iUt} & -\frac{\sqrt{2}\mu }{\kappa }%
\end{array}%
\right)  \notag \\
&&\times (-\sqrt{2}\kappa ).
\end{eqnarray}%
The eigenvalues of $h_{1}$ are $U\pm 2\kappa $\ and $U$, which result in the
periodic dynamics with a period of $\pi /\kappa $.

On the other hand, there exists an effective invariant subspace, in which
matrix $h_{2}$\ reduces to

\begin{equation}
\tilde{h}_{2}=\left( 
\begin{array}{ccc}
U & -\sqrt{2}\kappa e^{-iUt} & 0 \\ 
-\sqrt{2}\kappa e^{iUt} & 0 & -\sqrt{2}\kappa e^{-iUt} \\ 
0 & -\sqrt{2}\kappa e^{iUt} & -U%
\end{array}%
\right) 
\end{equation}%
in the large $\mu $\ limit. We note that matrices $h_{1}$\ and $\tilde{h}_{2}
$\ have the connection%
\begin{equation}
\exp (-ih_{1}t)=K\mathcal{T}\exp (-i\int_{0}^{t}\tilde{h}_{2}(t^{\prime })%
\mathrm{d}t^{\prime }),
\end{equation}%
where $\mathcal{T}$\ is the time-order operator and%
\begin{equation}
K=\left( 
\begin{array}{ccc}
1 & 0 & 0 \\ 
0 & e^{-iUt} & 0 \\ 
0 & 0 & e^{-2iUt}%
\end{array}%
\right) ,
\end{equation}%
which accords with the local correspondence we proposed in the main text.
For finite $\mu $, the efficiency of this correspondence is determined by
the value of $\mu $. To demonstrate this point, we numerically compute the
time evolution of an initial state $|\phi (0)\rangle =c_{1,\uparrow }^{\dag
}c_{1,\downarrow }^{\dag }\left\vert 0\right\rangle $\ under the two
Hamiltonians $H_{1}$\ and $H_{2}$. We employ the fidelities, given by%
\begin{eqnarray}
F(t) &=&|\left\langle \phi (0)\right\vert \exp (iH_{1}t) K\mathcal{T}  \label{Ft} \\
&&\times \exp (-i\int_{0}^{t}H_{2}(t^{\prime })\mathrm{d}%
t^{\prime })|\phi (0)\rangle |^{2},  \notag
\end{eqnarray}%
and%
\begin{equation}
F_{\mathrm{E}}(t)=\left\vert \left\langle \phi (0)\right\vert \exp
(iH_{1}t)|\phi (0)\rangle \right\vert ^{2},  \label{F_Et}
\end{equation}%
and%
\begin{equation}
F_{\mathrm{\Phi }}(t)=\left\vert \left\langle \phi (0)\right\vert \exp
(-i\int_{0}^{t}H_{2}(t^{\prime })\mathrm{d}t^{\prime })|\phi (0)\rangle
\right\vert ^{2},  \label{F_Phit}
\end{equation}%
to characterize the similarity of the two evolved states for difference
values of $\mu $. We note that when taking $t=2\pi /U$, we have $K=1$, which
can simplify the calculation without losing generality.

\end{document}